# The *w*-index: A significant improvement of the *h*-index

Qiang Wu

*School of Management, University of Science and Technology of China, Hefei 230026, China*
*E-mail: qiangwu@ustc.edu.cn*

I propose a new measure, the *w*-index, as a particularly simple and useful way to assess the integrated impact of a researcher's work, especially his or her excellent papers. The *w*-index can be defined as follows: If *w* of a researcher's papers have at least 10*w* citations each and the other papers have fewer than 10(*w*+1) citations, his/her *w*-index is *w*. It is a significant improvement of the *h*-index.

**1. Introduction**

The *h*-index was first proposed in 2005 by Hirsch[1] to quantify an individual's scientific achievement. Because of its simplicity and comprehensibility, it quickly attracts immense amount of interests from the scientific community. A number of derived Hirsch-type indices, such as the *g* index[2], *h(2)* index[3], *a* index[4,5], *r* index[5], *ar* index[5], *m* quotient[1,6], *m* index[6], $h_w$ index[7], $h_m$ index[8] and so on, have been put forward and discussed in the literature. While each indicator has its own advantages, they generally require a large amount of information and therefore lose the inherent simplicity of the *h*-index. However, are these indicators better than the *h*-index? Bornmann et al. separated the *h* index and eight derived Hirsch-type variants into two categories: One type, including *h* index, *m* quotient, *g* index and *h(2)* index, describes the main productive core of an individual's scientific output and provides the number of papers in that core; The other type — *a* index, *m* index, *r* index, *ar* index and $h_w$ index — pictures the impact of the papers in the core. Regression analysis shows that the latter can predict peer assessments better than the former[6]. According to Lehmann et al., in terms of both accuracy and precision, the mean number of citations per paper is also a better way of measuring scientific quality than the *h*-index[9]. To improve the accuracy of the *h*-index, whilst keeping the simplicity of the *h*-index, I propose a new measure, the *w*-index, as a particularly clear and useful measure of the broad impact of a researcher's masterpieces.

**2. What is the *w*-index?**

If all the papers of a scientist are ranked in descending order of the number of citations they received, the *w*-index is the highest number of papers one has that have each received at least 10*w* or more citations. Hence the papers on ranks *w*+1, *w*+2, … have no more than 10(*w*+1) citations. Thus, the *w*-index can be defined as follows: If *w* of a researcher's papers have at least 10*w* citations each and the other papers have fewer than 10(*w*+1) citations, his/her *w*-index is *w*. This also implies that one hasn't achieved the higher level: *w*+1 of his/her papers have at least 10(*w*+1) citations each. For example, Stephen Hawking has a *w*-index of 24, meaning 24 of his papers have been cited at least 240 times each, and at the same time he does not have 25 papers cited at least 250 times each. With 41 papers cited at least 410 times each, Ed Witten of the Princeton Institute for Advanced Study has a *w*-index of 41, much higher than Stephen Hawking's 24. While the *w*-index appears to be very similar to the *h*-index, by more accurately reflecting the influence of a scientist's top papers or their representative work, the *w*-index is a significant improvement over the *h*-index. Certainly, on the occasion of not causing the ambiguity, the *w*-index may be called as the 10*h*-index. Furthermore, the *w*-index still maintains the greatest advantage of the *h*-index — simplicity and clearness which can be understood and popularized easily, and usually can be found very quickly in the databases such as ISI Web of Knowledge, Scopus and Google Scholar.

## 3. Empirical analysis
### 3.1. Evaluation of 20 astrophysicists

To examine the accuracy of the *w*-index, I analyze seven measures of the author quality: the mean number of citations per paper, the number of papers published, the number of citations, the *h*-index, the *a*-index, the *g*-index and the *w*-index. I select 20 astrophysicists and access their scientific impact. The original data were taken in May, 2008 from the Thomson ISI Web of Knowledge database. There are three criteria to select the astrophysicists: first, each of them has already published a series of papers on the famous international journals and has at least 50 periodical papers that can be found in the ISI Web of Knowledge database; secondly, they have been engaged in the astrophysics research for more than 20 years; thirdly, they are still actively pursuing astrophysics research now. Of course, in order to ensure more types of scholars can be selected, this paper maintains a certain degree of randomness. However, these 20 astrophysicists all satisfy the three criteria. Results are listed in the Table 1. For privacy I use only the relevant code to represent each astrophysicist.

Table 1 shows some differences among the seven methods. First of all, here, I analyze the situation of the top five astrophysicists judged by these seven methods. In this respect, the rank measured by the *p*-method is very different from any other methods. For example, SBD is assessed as the most competitive scientist by almost all methods, but by the *p*-method, SBD is ranked only fifth. Also, JEB ranks 2nd according to the *w*-index, implying that JEB should have a number of high impact papers. Other methods also assess JEB at the 3rd or 4th places, but the *p*-method makes JEB in the 9th position. Furthermore, the 3rd place of VMA via the *p*-method is obviously too high, because the mean number of citations per paper of VMA ranks at the 14th with a value of only 31.17, which is very lower than the average citations of 37.39. As for HDJ, the 15th place given by the *p*-method is clearly too low. Ranked at the 5th, the mean number of citations per paper of HDJ is 47.23, about 10 citations higher than the average. And the *h*-index also shows a strong consistency with the *p*-method when evaluating VMA and HDJ, indicating bias in both methods. Similarly, the evaluation of VMA and HDJ made by the *g*-index and *c*-method is also slightly biased. Only the *w*-index and *a*-index give the relatively accurate results of VMA and HDJ, reflecting the different features that concern the average citation rates of these two researchers, and maintain a degree of consistency with the *m*-method. Certainly, the *m*-method also has its own weaknesses: in general, when appraising the impact of an individual's scientific work, sometimes, it is disadvantageous for those who have published a large number of papers while favorable for those who have issued fewer articles. In this case, aiming for the places of FEL and LJL, the *m*-method causes some conflicts with the other methods. The reason is that: LJL is a scholar who has published the most number of papers among all 20 astrophysicists, causing the judgment is somewhat low; and the number of papers issued by FEL is less, leading a high place.

Furthermore, after comparing the results from all methods, it is apparent that SML, DJG, DSA, SNS and SSN are at the bottom of the ranking. However, the $c$-method evaluates DJG at a slightly higher value, and the positions of DJG and SSN according to the $p$-method, at second and seventh places respectively, show greater inconsistency with other methods.

After evaluating all these methods, it is reasonable to believe in general the $p$-method is ineffective. Lehmann et al. made a similar judgment, and this is the reason why the $p$-method, being equivalent to papers/year method[9,11], is hardly used as an indicator of scientific quality.

Table 1 Results of the 20 astrophysicists

| code | $w$-index value | rank | $h$-index value | rank | $a$-index value | rank | $g$-index value | rank | $m$-method value | rank | $c$-method value | rank | $p$-method value | rank |
|---|---|---|---|---|---|---|---|---|---|---|---|---|---|---|
| SBD | 16 | 1 | 60 | 1 | 165.85 | 1 | 112 | 1 | 65.21 | 1 | 13890 | 1 | 213 | 5 |
| JEB | 14 | 2 | 50 | 3 | 111.46 | 4 | 78 | 4 | 52.86 | 3 | 7294 | 4 | 138 | 9 |
| LJL | 12 | 3 | 56 | 2 | 118.02 | 3 | 91 | 2 | 40.78 | 9 | 11663 | 2 | 286 | 1 |
| HDJ | 12 | 3 | 35 | 12 | 101.29 | 5 | 64 | 10 | 47.23 | 5 | 4297 | 10 | 91 | 15 |
| BRC | 11 | 5 | 41 | 7 | 156.46 | 2 | 87 | 3 | 54.23 | 2 | 8135 | 3 | 150 | 6 |
| KGA | 10 | 6 | 40 | 9 | 92.30 | 6 | 66 | 7 | 42.62 | 6 | 4944 | 8 | 116 | 12 |
| CCR | 10 | 6 | 46 | 5 | 90.57 | 7 | 71 | 5 | 41.88 | 7 | 5905 | 7 | 141 | 8 |
| GJE | 10 | 6 | 41 | 7 | 85.68 | 10 | 66 | 7 | 28.27 | 15 | 6135 | 6 | 217 | 4 |
| FEL | 9 | 9 | 30 | 15 | 85.87 | 9 | 54 | 14 | 50.10 | 4 | 2956 | 15 | 59 | 18 |
| CJP | 9 | 9 | 33 | 13 | 85.94 | 8 | 58 | 11 | 35.06 | 10 | 3576 | 14 | 102 | 14 |
| ATR | 9 | 9 | 36 | 11 | 76.61 | 13 | 57 | 12 | 31.23 | 13 | 3810 | 12 | 122 | 10 |
| VMA | 9 | 9 | 48 | 4 | 84.04 | 11 | 70 | 6 | 31.17 | 14 | 6989 | 5 | 220 | 3 |
| MJE | 9 | 9 | 45 | 6 | 79.41 | 12 | 65 | 9 | 41.53 | 8 | 4942 | 9 | 119 | 11 |
| MHL | 8 | 14 | 33 | 13 | 67.03 | 15 | 50 | 15 | 31.93 | 12 | 2714 | 16 | 85 | 16 |
| RRA | 8 | 14 | 38 | 10 | 69.63 | 14 | 56 | 13 | 34.17 | 11 | 3759 | 13 | 110 | 13 |
| SML | 6 | 16 | 22 | 19 | 54.50 | 17 | 38 | 19 | 23.88 | 18 | 1576 | 19 | 66 | 17 |
| DJG | 6 | 16 | 30 | 15 | 55.50 | 16 | 45 | 16 | 16.16 | 20 | 3895 | 11 | 241 | 2 |
| DSA | 5 | 18 | 24 | 18 | 51.08 | 18 | 39 | 18 | 25.71 | 16 | 1594 | 18 | 52 | 20 |
| SNS | 5 | 18 | 22 | 19 | 49.91 | 20 | 35 | 20 | 24.31 | 17 | 1313 | 20 | 54 | 19 |
| SSN | 5 | 18 | 29 | 17 | 51.00 | 19 | 42 | 17 | 17.51 | 19 | 2504 | 17 | 143 | 7 |

Notes: $m$-method — the mean number of citations per paper; $c$-method — the number of citations; $p$-method — the number of papers published.

Next, to test the relations among these methods, a rank Spearman correlation analysis is employed and the results are presented in Table 2. The highest correlation coefficient is found between the $g$-index and the $c$-method (0.967), because both methods rely heavily on citations. High correlation is also found between the $w$-index and the $a$-index (0.965), because they focus only on citations of outstanding papers. The $g$-index also shows a high correlation with the $h$-index (0.953). The minimum correlation is discovered between the $m$-method and the $p$-method (0.060), which uses the opposite methodology to deal with the number of papers published. The other two lower correlation coefficients are found between the $a$-index and the $p$-method (0.420) and between the $w$-index and the $p$-method (0.450). Thus, from the correlation coefficients, the seven measures can be approximately divided into two categories:

one category is fairly apart from the *p*-method, including *m*-method, *a*-index and *w*-index, and the correlations between them and the *p*-method are not significant at the 0.01 level; the other is relatively closer to the *p*-method, including *g*-index, *c*-index and *h*-index, and the correlations between them and the *p*-method are significant at the 0.01 level. In general, results close to the *p*-method are unreliable and indicate that they measure industry rather than ability[9]. Some experts have respectively demonstrated that the accuracy of the *m*-method and *a*-index is significantly higher than the *h*-index[6,9]. Therefore, the *w*-index is indeed an excellent method, because it is relatively far away from the *p*-method and obtains results that are similar to those obtained with the *a*-index and *m*-method.

Table 2 Spearman's correlation coefficients

|   | *w*-index | *h*-index | *a*-index | *g*-index | *m*-method | *c*-method | *p*-method |
|---|---|---|---|---|---|---|---|
| *w*-index | 1.000 | | | | | | |
| *h*-index | 0.827** | 1.000 | | | | | |
| *a*-index | 0.965** | 0.781** | 1.000 | | | | |
| *g*-index | 0.916** | 0.953** | 0.912** | 1.000 | | | |
| *m*-method | 0.829** | 0.631** | 0.877** | 0.735** | 1.000 | | |
| *c*-method | 0.873** | 0.940** | 0.850** | 0.967** | 0.624** | 1.000 | |
| *p*-method | 0.450 | 0.672** | 0.420 | 0.650** | 0.060 | 0.768** | 1.000 |

**Significant at 1%.

In summary, when characterizing the scientific impact of the 20 astrophysicists, the *w*-index and *a*-index are the most useful methods, and their results are consistent with each other; the *m*-method, *g*-index, *c*-index and *h*-index are also fairly effective, but with certain levels of bias; the *p*-method is not a good measure, diverging heavily from the other methods. The effectiveness of the *w*-index indeed surpasses that of the *h*-index. Of course, besides effectiveness, there are some other standards to qualify an index, such as the time needed to collect information, the clarity of understanding, and so on. In fact, compared with the other methods except the *w*-index, the *h*-index has indeed unique features: simplicity to be understood easily and saving computing time. Manuel Cardona commented that the *h-i*ndex can be obtained in about 30 seconds from the ISI Web of Knowledge[10]. If so, the *w*-index can be completed within 10 seconds, because it uses much less information than the *h*-index. Especially when appraising the scientists who have published numerous articles, the *w*-index can count only their most representative work and display its advantages better. The *w*-index also has the advantage over the *a*-index. It maintains the superiority of using less information, because the first step of the *a*-index is to find the *h*-index, and then it must precisely collect the specific number of citations of these *h* papers, needing more information than the *h*-index. Therefore, depending on less information and more exact explanation, the *w*-index is better that the *a*-index, and the meaning of the *w*-index is clear at a glance, just as the *h*-index. If your *h*-index is 10, it means that 10 of your papers that have each received at least 10 citations; if your *w*-index is 10, it denotes that 10 of your papers that have each received at least 100 citations. And the *a*-index does not obtain a simple definite concept so far.

### 3.2. The *w*-index of a few highest-ranked physical scientists

Table 3 Some of the highest-ranked scientists' *w*-index

| | Physicists, by *w*-index | |
|---|---|---|
| 1 | Ed Witten | 41 |
| | *Institute for Advanced Study, Princeton* | |
| 2 | Philip Anderson | 26 |
| | *Princeton University* | |
| 3 | Stephen Hawking | 24 |
| | *University of Cambridge* | |
| 4 | Frank Wilczek | 23 |
| | *Massachusetts Institute of Technology* | |
| 4 | Marvin Cohen | 23 |
| | *University of California, Berkeley* | |
| | Chemists, by *w*-index | |
| 1 | Kurt Wüthrich | 30 |
| | *Swiss Federal Institute of Biology, Zurich* | |
| 2 | Martin Karplus | 29 |
| | *Harvard University* | |
| 3 | George Whitesides | 28 |
| | *Harvard University* | |
| 4 | Alan Heeger | 26 |
| | *University of California, Santa Barbara* | |
| 5 | Elias James Corey | 24 |
| | *Harvard University* | |

To explain the significance of the *w*-index, I give another interesting example, further indicating that the index *w* is a superb and stable estimator of advanced scientific achievement. In 2007, some physical scientists with high *h*s are listed in *Nature*[12]: the physicists are Ed Witten, Marvin Cohen, Philip Anderson, Manuel Cardona and Frank Wilczek; the chemists are George Whitesides, Elias James Corey, Martin Karplus, Alan Heeger and Kurt Wüthrich; and the Stephen Hawking whose *h*-index is lower than those of them. Here, I present the *w*-indices of 10 scientists of them in Table 3. The original data were taken in May, 2008 from the ISI Web of Knowledge database. Compared with the *h*-index, the positions of scientists show an obvious change, which can not be explained by the different evaluation period of time. Within the physicists, Ed Witten remains among the best performer, at first position overall, and greatly exceeds other scientists. Philip Anderson ascends the second place, with Stephen Hawking following closely at third who is away from the rankings of the *h*-index. Moreover, formerly at the fourth place, Manuel Cardona's *w*-index is 20 and excluded from the new rankings, but his *h*-index is much higher than Stephen Hawking's[12]. The change of the chemists' place is also dramatic: Kurt Wüthrich jumps from previously fifth to first; George Whitesides droppes from earlier first to third. The new rankings of this paper based on the *w*-index, will focus on the quality of the outstanding scientists' representative articles, being closer to the reality, while the *h*-index pays more attention to the quantity of papers published. For example, according to the *w*-indices, 24 of Stephen Hawking's papers have been cited at least 240 times each, and at the same time, Manuel Cardona only has 20 papers have been cited at least 200 times each, indicating that Stephen Hawking is more influential because of his more significant papers. Once you use the $h$-index to quantify their scientific research output, the results will be completely different. In short, the $w$-index makes the experts to compete in the high-end displaying their papers' impact, and the $h$-index often lets the scholars take up the challenge in the mid-part.

Table 3 shows that the difference between physics and chemistry is slight and there appears to be certain similarity in the two types of science, but it remains to be seen whether or not this idea can be put into practice. Clearly, compared with the $h$-index, the $w$-index fills in the gaps in different fields of science, because from the $h$-indices listed in *Nature*[12] we can confirm that their values in chemistry tend to be higher than those in physics and the difference seems to be much larger at the high end than on average[1]. Aiming at the $w$-index, however, it is difficult to form such opinion since the averages of $w$-indices in chemistry and in physics are actually 27.4. Is the $w$-index really has the characteristic that can greatly narrow the differences of evaluation of some subjects, especially when they all belong to the field of physical sciences? When putting forward the $h$-index, Hirsch thought the differences of citation patterns among various subjects are so distinct that each field would need dissimilar thresholds[13]. But with the $w$-index, different subjects may indeed have similar citation models, so each field would need similar thresholds.

## 4. Summary and discussion

My preliminary research shows that the average of $h$-indices is about four times the mean value of the $w$-indices, i.e. $h \approx 4w$. Therefore, based on the Hirsch's conclusions[1] and this paper's results, I propose a number of hypotheses:

i) A $w$-index of 1 or 2, characterizes a researcher who has learned the rudiments of a subject.

ii) A $w$-index of 3 or 4, characterizes a scientist who has fully mastered the art of scientific activity.

iii) A $w$-index of 5, characterizes a successful researcher.

iv) A $w$-index of 10, characterizes outstanding individuals.

v) A $w$-index of 15 after 20 years of scientific activity, or 20 after 30 years, characterizes top scientists.

In fact, these forecasts have been somewhat validated from the $w$-indices of ten highest-ranked scientists, their $w$s all exceeding 20, and twenty successful astrophysicists, their $w$s all reaching 5. Of course, it is impossible to measure the entire ability of a scientist by using only one index. As the $w$-index depends on the citations, the potential problems such as the lag of being cited, discrepancy of databases, influence of namesakes and so on will limit the scope for its application. However, whether or not you believe it, the $w$-index will have a niche in the evaluation of scientific achievement.

Currently, the $h$-index has been used for evaluation of scientists[14,15,16], journals[17,18,19], conferences[20], scientific topics[21], research institutions[22,23], and so on. Similarly, in terms of both simplicity and accuracy, the $w$-index does have a bright prospect and display its prowess fully in these fields to measure the real impact of their representative articles.

**Acknowledgements**

I would like to express my gratitude to T. Fang for his helpful suggestions.